\documentclass[Letter]{aa}

\usepackage[T1]{fontenc}
\usepackage{pslatex}
\usepackage{amsmath}
\usepackage{amsfonts}
\usepackage{xcolor}
\usepackage{url}
\usepackage{bm}
\usepackage{graphicx}
\usepackage{amssymb}
\usepackage{soul}
\usepackage{booktabs}
\usepackage{array}
\usepackage[utf8]{inputenc}
\inputencoding{latin1}
\inputencoding{utf8}
\usepackage{appendix}

\def\lsim{ \lower .75ex\hbox{$\sim$} \llap{\raise .27ex \hbox{$<$}} }
\def\gsim{ \lower .75ex \hbox{$\sim$} \llap{\raise .27ex \hbox{$>$}} }

\newcommand{\bi}{\begin{itemize}}
\newcommand{\ei}{\end{itemize}}

\usepackage[colorlinks=true,linkcolor=blue,citecolor=blue]{hyperref}

\begin{document}

\title{Probing the low-energy particle content of blazar jets through MeV observations}
\titlerunning{Thermal bump in MeV blazars?}

\author{
F. Tavecchio\inst{1}
\and L. Nava\inst{1}
\and A. Sciaccaluga\inst{1,2}
\and P. Coppi\inst{3}
}
\authorrunning{Tavecchio et al.}

\institute{
INAF -- Osservatorio Astronomico di Brera, Via E. Bianchi 46, I-23807 Merate, Italy
\and
Dipartimento di Fisica, Universita` degli Studi di Genova, Via Dodecaneso 33, I-16146 Genova, Italy
\and
Department of Astronomy, Yale University, PO Box 208101, New Haven, CT 06520-8101, USA
}
\date{}

\voffset-0.4in



\abstract{Many of the blazars observed by {\it Fermi} actually have the peak of their time-averaged gamma-ray emission outside the  $\sim$ GeV {\it Fermi} energy range, at $\sim$ MeV energies.  The detailed shape of the emission spectrum around the $\sim$ MeV peak places important constraints on acceleration and radiation mechanisms in the blazar jet and may not be the simple broken power law obtained by extrapolating from the observed X-ray and GeV gamma-ray spectra. In particular, state-of-the-art simulations of particle acceleration by shocks show that a significant fraction (possibly up to $\approx 90\%$)  of the available energy may go into bulk, quasi-thermal heating of the plasma crossing the shock rather than producing a non-thermal power law tail. Other ``gentler" but possibly more pervasive acceleration mechanisms such as shear acceleration at the jet boundary may result in a further build-up of the low-energy ($\gamma \lesssim 10^{2}$) electron/positron population in the jet.  As already discussed for the case of gamma-ray bursts, the presence of a low-energy, Maxwellian-like ``bump'' in the jet particle energy distribution can strongly affect the spectrum of the emitted radiation, e.g., producing an excess over the emission expected from a power-law extrapolation of a blazar's GeV-TeV spectrum.   We explore the potential detectability of the spectral component ascribable to a hot, quasi-thermal population of electrons in the high-energy emission of flat-spectrum radio quasars (FSRQ). We show that for typical FSRQ physical parameters, the expected spectral signature is located at $\sim$ MeV energies. For the brightest {\it Fermi} FSRQ sources, the presence of such a component will be constrained by the upcoming MeV Compton Spectrometer and Imager (COSI) satellite.
}


\maketitle
\boldsymbol{}

\section{Introduction}

Even after decades of effort, a detailed understanding of the physical processes responsible for the phenomenology of extragalactic relativistic jets still eludes us (see, e.g., \citealt{Blandford19}). Basic questions related to jet dynamics, composition, and the role of magnetic fields await definitive answers. In particular, the nature of the mechanisms behind the dissipation of a jet's bulk outflow energy, be it initially in the form of particles or Poynting flux, and the subsequent acceleration of particles to ultrarelativistic energies, as evidenced by emission of some jets in the TeV band, remains a central problem (\citealt{Sironi15a,matthews20}). The two main competitors are diffusive shock acceleration (DSA) and magnetic reconnection (MR). In fact, the two mechanisms are somewhat complementary, since DSA can efficiently work only for flows with small magnetization ($\sigma$), while MR naturally requires high $\sigma$ (\citealt{Sironi15b}). Highly magnetized jets seem to be the natural outcome of launching mechanisms involving the interplay of magnetic fields and BH rotation (e.g. \citealt{Tchekhovskoy2011}), thus favoring MR. However, models of the emission observed from blazars indicate small magnetizations in connection with the emitting region(s) (\citealt{sikora2005,CelottiGhisellini08,TavecchioGhisellini16}), potentially supporting DSA. Recent results in the polarimetric channel by the IXPE satellite seem also to point to shocks as the main actors in the acceleration (e.g. \citealt{liodakis22}), although other interpretations are possible, including Poynting-dominated jets (e.g. \citealt{Bolis24}).


Particle-in-cell (PIC) simulations allow us to study in detail (albeit on small temporal and spatial scales) the acceleration processes, both in the case of DSA (e.g. \citealt{Sironi11,Sironi13,Crumley19,Groseli24}) and MR (e.g. \citealt{Sironi14,werner2018,petropoulou19,werner2024}). An established prediction for DSA is that, downstream of the shock, particles are heated and form a Maxwellian-like distribution, while only a small fraction ($\sim$ a few percent) of the particles, repeatedly  crossing the shock,  undergo DSA  and form a power law tail containing $\lesssim 10\%$ of the energy dissipated at the shock (e.g. \citealt{Spitk2008}). This kind of distribution is also derived through simulations adopting the Monte Carlo approach (e.g. \citealt{Summerlin12}). The presence of the prominent thermal bump should imprint clear spectral signatures (e.g. \citealt{giannios09}). The possible presence of these signatures has been in particular discussed for gamma-ray burst (GRB) \citep{eichler05,giannios09,warren18,gao24}, but conclusive observational evidence for a thermal component is lacking. On the other hand, MR is expected to produce smooth, power law-like spectra (e.g. \citealt{petropoulou19}). In this case, therefore, one does not expect any narrow, thermal component in the observed spectra. 

Our aim is to explore the potential signatures of the electron thermal bump in the emission of flat-spectrum radio quasars (FSRQ). These are powerful blazars characterized by a dominant $\gamma$-ray component, probably produced through the inverse Compton scattering (IC) of ambient radiation by relativistic leptons in the jet (e.g. \citealt{sikora94,ghisellinitavecchio09}). A back-of-the-envelope calculation suggests that, for typical parameters, electrons belonging to the thermal component emit in the MeV band, which corresponds to the maximum of the high-energy component of FSRQ (e.g. \citealt{sikora02,ghisellini17,marcotulli22}).\footnote{\cite{baring17} performed a similar study, but they assumed a cold thermal component, whose emission peaks in the soft X-ray band.} For several of these sources the observed flux in the MeV is within the reach of the upcoming Compton Spectrometer and Imager (COSI) satellite (\citealt{tomsick22}), giving the opportunity to test our scenario in the near future.

The paper is organized as follows: in Sect. 2 we present the model, in Sect. 3 we discuss the application to FSRQ and we present the results. Finally, in Sect. 4 we discuss the results and the observational prospects.

Throughout the paper, the following cosmological parameters are assumed: $H_0=70{\rm\;km\;s}^{-1}{\rm\; Mpc}^{-1}$, $\Omega_{\rm M}=0.3$, $\Omega_{\Lambda}=0.7$.

\section{The model}
\label{sec:model}

As an example calculation of what one might see if a significant fraction of a jet's dissipated energy goes into bulk heating rather than non-thermal acceleration, we follow a standard one-zone approach, e.g., \cite{MaraschiTavecchio2003}. The region where the dissipation occurs is modeled as a sphere with (comoving) radius $R$, moving with bulk Lorentz factor $\Gamma$ at an angle $\theta_{\rm v}$ with respect to the line of sight\footnote{Unless noted, all physical parameters, except $\Gamma$ and $\theta_{\rm v}$, are measured in the jet frame.}. The region is filled with a tangled magnetic field of strength $B$. Relativistic electrons emit through synchrotron and IC mechanisms. For the IC targets, we consider both internally produced synchrotron radiation (synchrotron self-Compton, SSC) and an external component (external Compton, EC) dominated by the quasar's broad-line region (BLR). The BLR spectrum is a black body with (observer frame) temperature $T$ and  energy density $U_{\rm ext}.$ . 

We do not attempt to model the dissipation/acceleration process. Rather, we assume the rapid production (``injection") of energetic electrons (or pairs), with an initial hybrid, thermal plus non-thermal (power law) energy distribution following \cite{giannios09}:
\begin{equation}
\label{eq:injection}
Q(\gamma) = 
\begin{cases}
K_eQ_{\rm th}(\gamma)e^{-\gamma_{\rm nth}/\gamma_{\rm c}} & {\rm if}\quad \gamma< \gamma_{\rm nth}\\
K_eQ_{\rm th}(\gamma_{\rm nth})\left(\frac{\gamma}{\gamma_{\rm nth}}\right)^{-p}e^{-\gamma/\gamma_{\rm c}} & {\rm if}\quad \gamma> \gamma_{\rm nth}
\end{cases}
\end{equation}
where $K_e$ is a normalization and the thermal distribution is given by the Maxwell-Juttner distribution:
\begin{equation}
    Q_{\rm th}(\gamma)=\frac{\beta \gamma^2}{\gamma_{\rm th}K_2(1/\gamma_{\rm th})}e^{-\frac{\gamma}{\gamma_{\rm th}}},
\end{equation}
with $K_2$ the modified Bessel function of the second kind. The parameter $\gamma_{\rm th}$ plays the role of an effective temperature $T$ of the quasi-thermal distribution ($\gamma_{\rm th}=kT/m_e c^2$, with $k$ the Boltzmann constant and $m_e$ the electron rest mass), while  $\gamma_{\rm nth}$ in Eq.\ref{eq:injection} is the minimum Lorentz factor of the non-thermal tail, described as a power law with slope $p$ with cut-off at $\gamma_{\rm c}$ (with $\gamma_{\rm c} \gg \gamma_{\rm nth}$). These ``injected'' particles then cool and radiate.

The jet rest-frame total injected luminosity of this hybrid (thermal and non-thermal) electron population is:
\begin{equation}
    L_e= m_ec^2 V \int_1^{\infty} \gamma Q(\gamma)d\gamma,
\end{equation}
where $V$ is the volume of the emitting region.
It is useful to introduce the fraction of energy contained in the non-thermal tail with respect to the total one:
\begin{equation}
    \delta\equiv\frac{\int_{\gamma_{\rm nth}}^{\infty}\gamma Q(\gamma)d\gamma}{\int_1^{\infty} \gamma Q(\gamma)d\gamma}.
\end{equation}

We also define the average Lorentz factor of the electron population:
\begin{equation}
   \langle\gamma\rangle =\frac{\int_1^{\infty}\gamma Q(\gamma)d\gamma}{\int_1^{\infty} Q(\gamma)d\gamma}.
\end{equation}

In the standard internal shock model the energy of the electrons in the post-shock region comes from the randomization of the bulk kinetic flux of the incoming protons.
It is customary to define the parameter $\epsilon_e$ as the fraction of the energy available from the shock that is conveyed to {\it non-thermal} electrons (e.g., \citealt{sari98}). In our case this can be expressed as:
\begin{equation}
\epsilon_e=\frac{m_ec^2\int_{\gamma_{\rm nth}}^{\infty}\gamma Q(\gamma)d\gamma}{\dot{n}_p m_p c^2 (\Gamma_{\rm u} -1)},
\end{equation}
where $\Gamma_{\rm u}$ is the Lorentz factor of the upstream fluid  in the downstream frame, $m_p$ is the proton mass, $\dot{n}_p$ is the number flux of the protons (as measured in the downstream frame).
Using the parameters introduced above it is possible to express $\epsilon_e$ as:
\begin{equation}
\epsilon_e=\frac{m_e}{m_p} \eta_{\pm} \frac{\langle \gamma \rangle \delta}{\Gamma_{\rm u} -1}.
\label{eq:epsilone}
\end{equation}
Here we have defined the multiplicity $\eta_{\pm}=\dot{n}_e/\dot{n}_p$, where $\dot{n}_e=\int Q(\gamma) d\gamma$ is the total injection rate of the electrons (both thermal and non-thermal). In a normal $ep$ plasma, $\eta_{\pm}=1$. We will also explore the case of a pair-enriched jet, for which $\eta_{\pm}>1$ (e.g. \citealt{SikoraMadejski00}). In this case, protons share their energy with more than one lepton, causing $\langle\gamma\rangle$ to decrease for constant $\delta$ and $\epsilon_e$ (see Eq. \ref{eq:epsilone}). 

The injected energy distribution, Eq. \ref{eq:injection}, is used to calculate the electron energy distribution (EED) reached by the electrons after a light crossing time of the emission region, $N(\gamma)$, with the standard continuity equation (e.g. \citealt{chiabghis99}):
\begin{equation}
\begin{split}
\frac{\partial N(\gamma,t)}{\partial t}=\frac{\partial}{\partial \gamma}\left [ \dot{\gamma} \, N(\gamma,t)\right] + Q(\gamma),
\label{eq:ngamma}
\end{split}
\end{equation}
where $\dot{\gamma}=\dot{\gamma}_{\rm s} + \dot{\gamma}_{\rm IC}$ is the energy-dependent radiative cooling rate (including synchrotron and IC emission) of electrons with Lorentz factor $\gamma$. For simplicity, we neglect the adiabatic cooling of the particles and the escape from the source. Eq. \ref{eq:ngamma} is solved numerically using the robust implicit method of \cite{ChangCooper} to find the EED at time $t_{\rm em}=R/c$, for which we calculate the resulting emission (see e.g. \citealt{ghisellinitavecchio09}).

\begin{figure}
    \centering
    \hspace*{-0.3 truecm}
    \includegraphics[width=1.\linewidth]{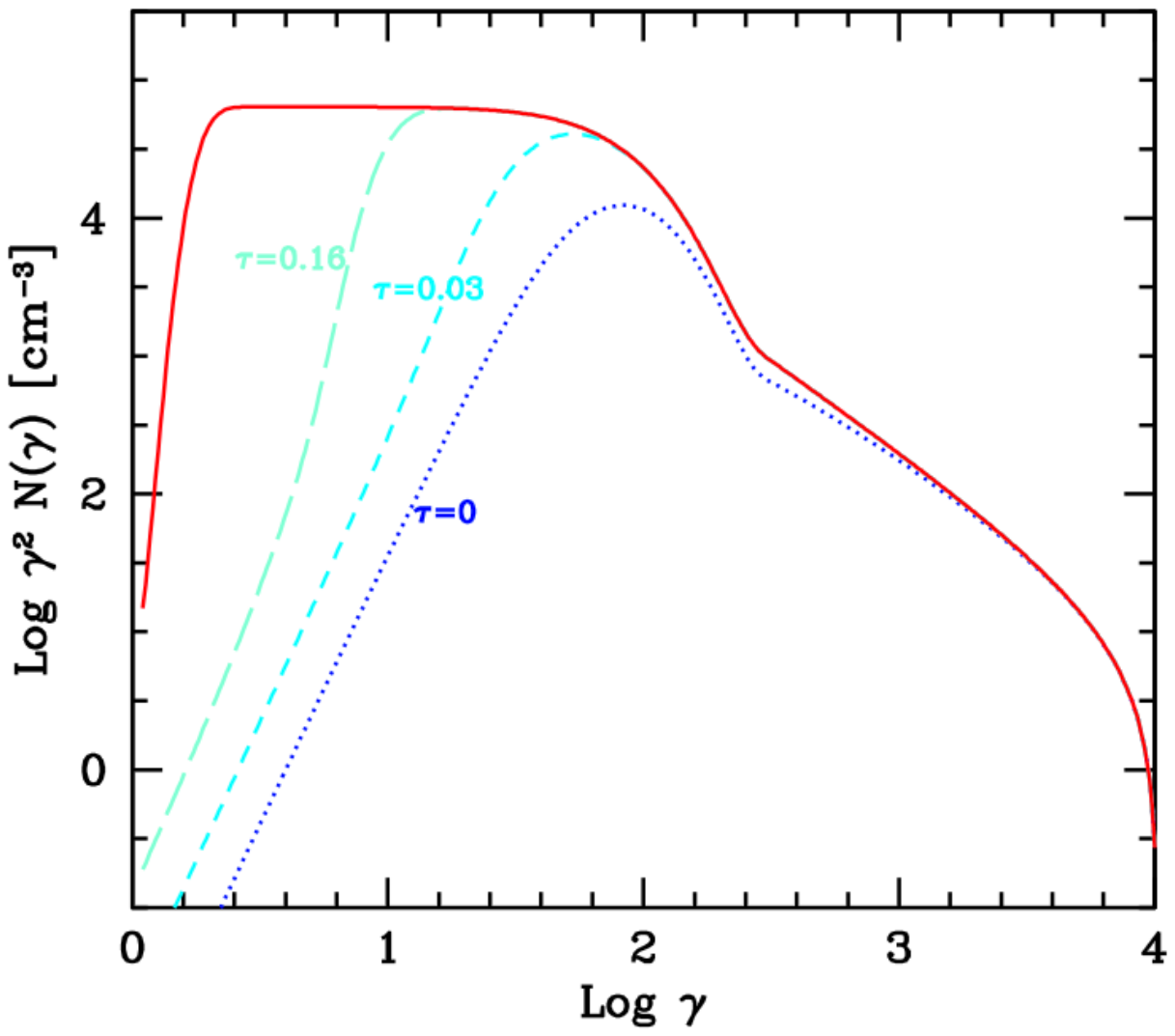}
    \caption{Electron energy distribution in the emission region calculated at different times (given in units of the light crossing time, $R/c$) with the parameters used for model A. The equilibrium distribution comprises a cooled low energy tail $N(\gamma)\propto \gamma^{-2}$, the high-energy tail (rapidly decreasing as a function of energy) of the thermal Maxwellian (which has the peak at $\gamma\simeq 100$) and the high-energy non-thermal power law with slope $p+1$ (with $p=2.3$ is the slope of the injected power law) up to the cut-off at $\gamma_c$.}
    \label{fig:dist}
\end{figure}

\subsection{Fixing parameters}

Since we intend to explore the effects of the complex EED on the spectral energy distribution (SED) of FSRQ, we assume benchmark values for the physical parameters of the blazar emission region inspired by modeling of FSRQ (e.g. \citealt{tavecchio00,Ghisellini10}). We assume $R=5\times 10^{16}$ cm, $B=1.6$ G, $\Gamma=20$, $\theta_{\rm v}=3.7$ deg (which combine to give a Doppler factor $D=15$). For the external radiation we assume that the dominant radiation field is that associated to the BLR, approximated (in the source rest frame) as a black body peaking at $\nu_{\rm ext}=2\times 10^{15}$ Hz (see the discussion in \citealt{tavecchio08}) with energy density $U_{\rm ext}=2.5\times 10^{-2}$ erg cm$^{-3}$ (e.g. \citealt{ghisellini08}).
We further assume that the source is located at the typical redshift $z=2$.

Besides the main physical quantities related to the emission region, our model depends on the parameters of the injected EED, $Q(\gamma)$, which are uniquely fixed once $\delta$, $\langle \gamma \rangle$, $L_e$, $p$ and $\gamma_{\rm c}$ are specified.

A possible scenario for the origin of the shock at which electrons are heated and subsequently accelerated to high-energies is that invoking the interaction of different portions of the flow characterized by different speed (internal shocks, e.g. \citealt{Spada01}). In this case we expect a mildly relativistic shock, with $\Gamma_{\rm u}\sim 1.5-2$. These shocks can efficiently accelerate particles only for low magnetization of the plasma (\citealt{Sironi15b}). Indeed, FSRQ jets at the distance where the observed radiation is produced is expected to have a small magnetization (e.g. \citealt{sikora2005,CelottiGhisellini08}). In these conditions we can rely on the results of \cite{Crumley19}, that report the analysis of PIC simulations for subluminal, high-Mach number, mildly relativistic ($\Gamma_{\rm u}=1.7$) weakly magnetized jets. The simulations show the development of a thermal component with average Lorentz factor around $\langle \gamma \rangle \approx 300$, and of a non-thermal power law with slope 2.2-2.3 containing a small fraction of the shock energy, $\epsilon_e\approx 10^{-3}$. However this last result could depend on the detailed configuration (i.e. inclination of the magnetic field and magnetization).  
PIC simulations of weakly magnetized, relativistic ($\Gamma_{\rm u}>3$) shocks show that the fraction of energy conveyed to non-thermal electrons can be much larger than that estimated in the mildly relativistic case, of the order of $\epsilon_e\approx 10^{-1}$ (\citealt{Sironi11}). These conditions could apply to FSRQ if the acceleration occurs at a stationary oblique recollimation shock (e.g. \citealt{BodoTavecchio18,zech21}). 

Given the large uncertainties related to the physical set-up of the emission region and the physical parameters related to the acceleration process, 
for definiteness in the following we will consider a benchmark model (model A hereafter) with fixed $\delta=0.15$, $\langle \gamma \rangle = 100$ and $p=2.3$. With this parameters $\epsilon_e\simeq 10^{-2}$ for a $ep$ plasma, in line with the results of the simulations mentioned above. However, we will also explore and discuss the effect of different values of the parameters on the resulting spectrum. In all models we  fix $\gamma_{\rm c}=8\times 10^4$. The value of $\gamma_{\rm c}$ is relatively unimportant (as long as $\gamma_{\rm c}\gg 10^3$), since the high-energy end of the EC component is mainly shaped by the effect of the KN cross section (\citealt{tavecchio08}).

\section{Results}
\label{sec:results}

The strong radiative losses (dominated by EC), imply that the system is in fast-cooling regime. The injected electrons belonging to the thermal bump quickly cool, forming a power law $\propto \gamma^{-2}$ down to very low Lorentz factors ($\gamma \approx 2$, see Fig. \ref{fig:dist} for which we assume the parameters of model A, see Table \ref{table:param}). Therefore, the EED reached at $t_{\rm em}=R/c$, is composed by this power law up to the peak of the relativistic Maxwellian, followed by a rapid decline in correspondence of the exponential part of the thermal peak and, finally, by the (cooled) non-thermal power law (with slope $p+1$) above $\gamma_{\rm nth}$ up to $\gamma_{\rm c}$.

\begin{table}
\caption{Parameters of the models}
\centering
\begin{tabular}{ccccc|c}
\hline
Model  & $L_e$ &$\delta$ & $\langle \gamma \rangle$ & $\eta_{\pm}$ & $\epsilon_e$ \\
\hline
A & 8.3 & 0.15 & 100 & 1& $1.7\times 10^{-2}$  \\
B & 8.0 & 0.05 & 100 &  1& $6\times 10^{-3}$\\
C & 4.2 & 0.15 & 50 & 2 & $1.7\times 10^{-2}$ \\
D & 20.2 & 0.15& 300 & 1 & $4.4\times 10^{-2}$\\
E & 8.4 & 0.30& 100 & 1 & $3\times 10^{-2}$\\
\hline
\end{tabular}
\tablefoot{
The last row reports $\epsilon_e$ derived from the input parameters with Eq. \ref{eq:epsilone} and $\Gamma_{\rm u}=1.5$. The injected electron luminosity $L_e$ is in units of $10^{44}$ erg s$^{-1}$.}
\label{table:param}
\end{table}

\begin{figure}
    \centering
    \hspace*{-0.45 truecm}
    \includegraphics[width=0.95\linewidth]{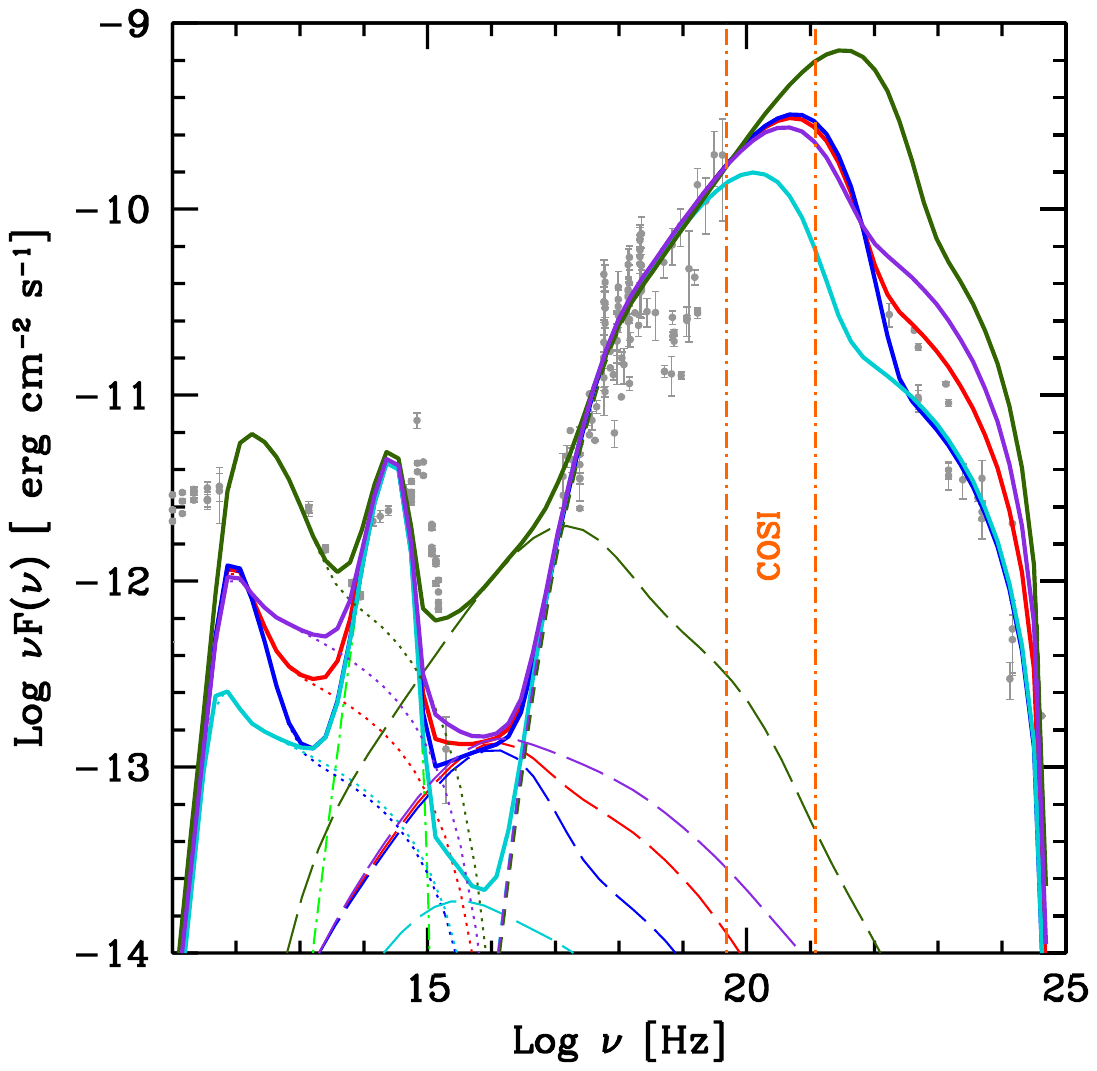}
    \vspace*{-2.25 truecm}
    \caption{SED calculated with the model described in the text. Colors refer to model A (red), B (blue), C (light blue), D (green) and E (violet). The dashed-dot light green line shows the contribution from the accretion disk. We also report the contribution from the synchrotron (dotted), SSC (long dashed) and EC (short dashed). The vertical orange lines show the COSI energy band. For reference, in gray we also show the observational datapoints of the FSRQ 0836+710 ($z=2.1$) from ASI-SSDC.
    }
    \label{fig:sed}
\end{figure}

The shape of the EED translates into a complex EC spectrum, with the notable presence of a ``bump" produced by the thermal electrons belonging to the Maxwellian (see Fig. \ref{fig:sed}). The expected observed energy of the peak can be easily estimated, since $E_{\rm p,obs}\approx h\nu_{\rm ext}\gamma^2_{\rm p} \Gamma D /(1+z)$, where $\gamma_{\rm p}\approx 80$ is the Lorentz factor at the the peak of the Maxwellian. We therefore find $E_{\rm p,obs}\simeq 5$ MeV.

The resulting observed SED is shown in Fig. \ref{fig:sed} (red). The shape of the EED is clearly recognized in the EC bump, whose maximum corresponds to the emission of the electrons at the thermal peak. Note also that in the synchrotron part of the SED the bump is partially visible but the self-absorption of the spectrum, effectively cutting the emission at frequencies below $10^{12}$ Hz, strongly limits the effect. Note also that the imprint of the EED in the SSC component is smoothed. It is therefore clear that the details of the EED can be probed only in the MeV band. Indeed, the extrapolation of the GeV spectrum to lower energies would clearly underpredict the flux in the MeV band. This {\it ``MeV excess"} would be the smoking gun of the presence of a thermal component in the EED. 

The other lines in Fig. \ref{fig:sed} show the SED for different values of $\langle \gamma \rangle$ and $\delta$ (see Table \ref{table:param}). For the sake of comparison, we fix the luminosity emitted in the X-ray band (acting on $L_e$) and we keep constant the parameters of the region ($B$, $R$, $\Gamma$, $\theta_{\rm v}$). 

Model B (blue lines) assumes a lower fraction of energy into the non-thermal tail ($\delta=0.05$) with respect to case A ($\delta=0.15$).  While the thermal component is similar to case A, producing the same peak in the SED, the lower $\delta$ determines a lower non-thermal tail in the GeV band. This case is therefore characterized by a larger ratio between the MeV and the GeV fluxes, making the discrepancy between the extrapolated GeV flux and the actual MeV flux even more severe.

In model C (light blue) we assume a pair-rich jet with multiplicity $\eta_{\pm}=2$ (i.e. one $e^{\pm}$ pair every two protons). In this case the energy shared by protons to leptons must be distributed among more particles, decreasing the average Lorentz factors of the population. Correspondingly, the effective temperature of the thermal component decreases, shifting the peak of the EC bump to lower energies. 

Model D is characterized by a larger $\langle \gamma \rangle=300$. The thermal peak shifts to high energies, exceeding 10 MeV. Due to the onset of the KN regime at energies above the EC peak, the transition between the thermal and the non-thermal part is smoother than in the other cases, making the identification of the two electron components more challenging. than in previous cases. Finally, case E (violet) is for a very high $\delta=0.3$. In this case the power law increases its level with respect to the Maxwellian component, determining a less pronounced bump in the SED.

\section{Discussion}
\label{sec:discussion}

Our results show that, within the large uncertainties 
related to the acceleration process and the composition of the jet, the peak of the thermal component is expected in the 0.1-10 MeV range, as observed for the most powerful FSRQ. Joint MeV and GeV observations can thus be exploited to trace the shape of the bump and potentially uncover the presence of the thermal component. We plan to perform dedicated simulations to assess the feasibility of observations with COSI \citep{tomsick22}.

In the alternative view invoking acceleration through MR, the anticipated spectrum is well approximated by simple power laws \citep{petropoulou19}. Therefore, in principle, the absence of the thermal bump could support the MR scenario. However, one should keep in mind that the prominence of the thermal component with respect to the non-thermal power law and, therefore, the possibility to disentangle the two  corresponding spectral components, is strictly related to the parameter $\delta$.  Our results suggest that the two components could be relatively easily identified even for cases with a large fraction of energy in non-thermal electrons, $\delta=0.3$. Recent PIC simulations \citep{Groseli24} suggest that larger $\delta$ ($>0.5$) can be reached in unmagnetized flows (associated to, e.g., GRB afterglows), but it is unclear if this result can be easily extended to (moderately) magnetized cases suitable for FSRQ. 

The strong anticorrelation between the peak energy of thermal component (and hence of the EC peak) and the pair content $\eta_\pm$, could already be used to rule out the case of a highly enriched plasma. In fact a high multiplicity would imply a peak below the MeV band, in contrast with observations. Unfortunately, the uncertainties on the physical parameters associated to the acceleration (in particular $\epsilon_e$) preclude definite conclusions. For instance, $\epsilon_e \sim 0.1$ (as derived for highly relativistic shocks) would allow $\eta_\pm \sim 20$ (as suggested, e.g., by \citealp{SikoraMadejski00}). In any case, a larger pair content seems to be excluded.

In this paper we have studied the case of the most powerful FSRQs, for which the thermal component naturally falls in the MeV band. For other types of blazars the situation is less straightforward. Less powerful FSRQ display the EC peak at higher energies (100 MeV-1 GeV, e.g. \citealt{marcotulli22}). If related to the thermal bump, such high peak energies would imply correspondingly high electron temperature, possibly related to larger dissipation parameter $\epsilon_e$. We remark that in this case, the high-energy part of the EC component would be affected by KN effects that smooth the spectrum, making difficult to identify the spectral structure (this effect is already visible in the D model in Fig. \ref{fig:sed}). At even lower power, blazars of the BL Lac type (LSP and HSP), display peaks above 10 GeV, which require electrons with $\gamma \gtrsim 10^4$ (e.g., \citealt{tavecchio10}), incompatible with thermal components from mildly relativistic shocks. In this case the thermal bump would instead appear in the low energy part of the IC component. We however remark that these sources produce high-energy radiation mainly through SSC, resulting in quite smooth spectra even for prominent thermal bumps (as also visible in Fig. \ref{fig:sed}).

An interesting issue concerns variability. Indeed, one can expect that the physical parameters controlling the EED ($\delta$, $\langle \gamma \rangle$) vary in time and determine the change of the position of the thermal bump and the relative level of thermal and non-thermal emission. The complex interplay between these effects could be in principle tracked by COSI (and {\it Fermi}) for the brightest FSRQ, with MeV flux exceeding $10^{-10}$ erg cm$^{-2}$ s$^{-1}$. We plan to study the feasibility of this approach in a future publication.


\begin{acknowledgements}
 We thank G. Ghisellini and E. Sobacchi for fruitful discussions. We acknowledge financial support from a INAF Theory Grant 2022 (PI F.~Tavecchio). This work has been funded by the European Union-Next Generation EU, PRIN 2022 RFF M4C21.1 (2022C9TNNX). Part of this work is based on archival data provided by the Space Science Data Center-ASI.
\end{acknowledgements}

\bibliographystyle{aa}
\bibliography{tavecchio}

\end{document}